\documentstyle[axodraw,aps,preprint]{revtex}

\begin{document}
\preprint{USC(NT)-97-09, nucl-th/9711053} 
\title{Chiral Perturbation Theory 
and Nuclear Physics \footnote{ 
Presented at the Georgia Theoretical Physics 
Symposium in honor of Don Robson's 
60th birthday} 
}

\author{F.Myhrer }

\address{Department of
Physics and Astronomy, 
University of South Carolina \\ 
Columbia, SC 29208, U. S. A.}

\maketitle

\vspace{-5mm} 

\begin{abstract}
Chiral symmetry serves as a guiding principle in 
low-energy hadron dynamics. 
An effective lagrangian, which explicitly 
breaks chiral symmetry via a small mass term, 
allows for a systematic method of calculating 
higher order corrections to tree-level results. 
This is applied to radiative muon capture on hydrogen, 
and to
proton-proton $\pi^0$ production near threshold. 
Both processes pose theoretical challenges in light of recent
experimental results. 
\end{abstract}

\narrowtext 


\section{Introduction} 

One of the more challenging research topics 
of modern nuclear physics is to relate the 
strong interaction dynamics of the quarks 
to the hadronic degrees of freedom 
of the nucleus. 
We know that the nucleons consists 
of at least three valence quarks, and we 
consider the quark dynamics to be 
determined by QCD. 
However, the predictions 
of QCD in hadronic low energy phenomena 
have proven too difficult to evaluate. 
For example, no one has been able to derive 
from QCD the strong nuclear forces between two nucleons. 
The analogous QED problem would be 
to calculate the van der Waals forces 
between two atoms, a task 
which has been accomplished successfully. 
Therefore, to investigate the 
low-energy strong hadronic interactions, 
the second best strategy would be to construct  
QCD inspired models 
(which all have their flaws), 
and/or to assume that the symmetry properties of QCD 
will manifest themselves also on the hadronic level. 
In this paper I will concentrate on 
hadronic manifestations and predictions of  the almost 
perfect chiral symmetry of the QCD lagrangian. 
Provided the quarks are massless 
the QCD lagrangian is chiral symmetric. 
In this ideal case the chirality (or helicity) of 
a quark is conserved, meaning
a right- and a left- handed quark 
will always retain their handedness. 
A quark mass term in the lagrangian will violate this symmetry 
and will allow left- and right- handed quarks to mix 
(or interact via spin-flip). 
Fortunately the u- and d- quark masses, 
$m_u \approx$ 7 MeV
and $m_d \approx$ 15 MeV, 
are small compared to the 
QCD scale $\Lambda_{QCD} \sim$ 200 MeV. 
It is therefore reasonable to 
consider the explicit breaking of chiral symmetry 
via the quark mass term in ${\cal L}_{QCD}$ as
a small perturbation. 
(In the following we shall not consider 
the implied isospin 
violations which we know is small, 
but insist that $m_u = m_d = m_q$.) 
From the Gell-Mann, Oakes and Renner (GOR) 
relation the pion mass squared: $m_\pi^2 \propto m_q$. 
This means
that the pions are massless in the chiral symmetric limit 
(when $m_q \ne 0$). 
Below we shall argue that we 
can calculate the corrections to the
chiral symmetry predictions of 
hadronic observables in a 
consistent, perturbative manner, via an approach called 
chiral perturbation theory (ChPT).

After a short presentation of the essential points 
of  ChPT, 
two examples of low energy hadronic physics will 
be presented where ChPT might give some valuable insight.
Finally  a brief overview of consequences of 
chiral symmetry considerations as applied to 
dense nuclear matter 
and the question of possible kaon condensation 
in neutron stars. 
It was this last question and its possible consequences 
in supernova explosion by Brown and Bethe \cite{bb94} that 
triggered our investigations of ChPT.

\section{Chiral symmetry and the effective lagrangian}

The seminal work of 
Gasser and Leutwyler \cite{gl84}
has shown that ChPT is a very powerful 
and successful technique for 
gaining more insight of hadronic phenomenology
at low energies.
Its application includes the physics of 
the pseudoscalar mesons ($\pi$,K,$\eta$)
and their interactions  \cite{ecker}, 
the pion-nucleon systems \cite{bkm95} 
and to some extent the 
few-nucleon systems \cite{vankolck,rho}. 

The effective chiral lagrangian of ChPT 
${\cal L}_{\rm{ch}}$ 
that describes the low energy 
($E < \Lambda_\chi \sim$ 1 GeV) 
hadronic phenomena  
involves an SU(2) matrix $U(x)$
which  is non-linearly related to the pion field, 
e.g., 
$U(x) = \sqrt{1-[\bbox{\pi}(x)/f_\pi]^2}
+i\bbox{\tau} \!\cdot\! \bbox{\pi}(x)/f_\pi $.\footnote{  
Another commonly used parameterization
is $U(x)=\exp[i\bbox{\tau} \!\cdot\! 
\bbox{\phi}(x)/f_\pi]$.  }
In the meson sector,
the sum of chiral-invariant monomials 
constructed from $U(x)$ and its derivatives 
constitute the chiral-symmetric part of
${\cal L}_{\rm{ch}}$.
The symmetry-breaking part of ${\cal L}_{\rm{ch}}$
is given via 
a mass matrix ${\cal M}$, 
the chiral transformation of which is
dictated by that of the quark mass term 
in the QCD lagrangian. 
${\cal L}_{\rm{ch}}$ is written as an expansion 
in powers of $\partial_\mu $ 
and the mass matrix ${\cal M}$, both 
characterized by a generic pion momentum 
$\tilde{Q} \ll  \Lambda_\chi$, 
where $\tilde{Q}$ 
represents either the pion momentum $q$  or 
the pion mass $m_\pi$. 
This means 
each term appearing in ${\cal L}_{\rm{ch}}$ 
carry a factor 
$(\tilde{Q}/\Lambda)^{\bar{\nu}}$, 
where the chiral order index $\bar{\nu}$ 
is defined by 
%
$\bar{\nu} \equiv d -2, $
%
where $d$ is the summed power of
the derivative and the pion mass involved in this term.
This suggests the possibility
of describing low-energy 
hadronic phenomena
in terms of ${\cal L}_{\rm{ch}}$
that contains only 
a manageably limited number 
of terms of  low chiral order. 
This is the basic idea of ChPT.

The heavy fermion formalism
(HFF)\cite{jm91} allows us 
to easily extend ChPT to the 
meson-nucleon system. 
This HFF is used 
since $\partial_0$ acting on the ordinary Dirac field $\psi$ 
describing the nucleon 
yields $\sim M $, a mass which is not small
compared with $\Lambda_\chi$.
In HFF,  $\psi$ 
is replaced by the heavy nucleon field $N(x)$ 
and the accompanying ``small  field"
$n(x)$ through the ``transformation" 
$\psi(x)
 = \exp(-i M v\cdot x)\,
[N(x) + n(x)] $. 
To obtain these new fields 
we use the 
projection operators 
${\cal P}^v_\pm$ = 
(1$\pm /\hskip -0.52em{v}) / 2$ 
to project out the ``heavy"-, $N$,  
 and ``small"-, 
$n$, fields from the 
Dirac spinor. 
By definition 
$N(x) \equiv {\cal P}_+^v exp(i M v \cdot x) \psi (x)$.  
%
%
Here the four-velocity $v_\mu$ is
assumed to be static, i.e.,
$v_\mu\sim (1,0,0,0)$. 
A systematic 
elimination of the ``small" 
field $n(x)$ in favor of $N(x)$
leads to an expansion in $\partial_\mu/M  $.
Since $M \approx$ 1 GeV $\approx\Lambda_\chi $,
an expansion in $\partial_\mu/M $ 
may be treated like 
the expansion in $\partial_\mu/\Lambda_\chi$. 
In HFF 
${\cal L}_{\rm{ch}}$ 
consists of chiral symmetric monomials 
constructed from $U(x)$, $N(x)$ and 
their derivatives in addition to 
the symmetry-breaking terms. 
Including the fermions the 
chiral order $\bar{\nu}$ in HFF 
is defined by 
$ \bar{\nu} \equiv d + n/2 -2$,
where $d$ is, as before,
the summed power of
the derivative and the pion mass, 
while $n$ is the number of nucleon fields 
involved in a given term  \cite{wei90}.
As before, each  term in ${\cal L}_{\rm{ch}}$ 
carry a factor 
$(\tilde{Q}/\Lambda)^{\bar{\nu}} \ll 1$. 
We use
${\cal L}_{\mathrm{ch}}$ 
as given in \cite{bkm95}.

In the most general form
the effective lagrangian ${\cal L}_{\mathrm{ch}}$ 
involving pions and heavy nucleons
in external weak- and electromagnetic-fields 
consistent with chiral symmetry,  
is written in increasing chiral order as: 
\begin{equation}
{\cal L}_{\mathrm{ch}} = 
{\cal L}_{\pi}^{(0)} + 
{\cal L}_{\pi N}^{(0)} + {\cal L}_{\pi N}^{(1)} +\cdots,
\label{Lag} 
\end{equation}
Here ${\cal L}^{(\bar{\nu})}$ represents
terms of chiral order $\bar{\nu}$, and  
the explicit expressions 
for the ${\cal L}_{\pi}^{(0)}$, 
${\cal L}_{\pi N}^{(0)}$ and ${\cal L}_{\pi N}^{(1)}$,
which only include terms of direct relevance for our calculations, 
are 
\begin{eqnarray}
{\cal L}_{\pi}^{(0)} \, &=& \, 
\frac{f_\pi^2}{4} {\mathrm{Tr}} \left [
D_\mu U D^\mu U \right ] \,
 + m_\pi^2 (U^\dagger +  U - 2) ] \,  + \dots 
\label{lpi0} \\
{\cal L}_{\pi N}^{(0)} \, &=& \,  
{\bar N} \left \{  i v\cdot D + 
g_A S\cdot u \right \}  N 
 - \frac12 \displaystyle \sum_A 
      C_A (\bar{N} \Gamma_A N)^2 
\label{lpin0} \\
{\cal L}_{\pi N}^{(1)} \, &=& \, 
{\bar N} \Big \{ \frac{1}{2M} 
(v\cdot D)^2 - \frac{1}{2M} D\cdot D
- \frac{i g_A}{2 M} \{ S\cdot D , v\cdot u \}_{+} 
\nonumber \\ 
 & + &  2c_1 m_\pi^2 \bar{N} N \mbox{Tr} 
( U + U^\dagger - 2 ) 
  +  (c_2 \!-\! \frac{g_A^2}{8 M }) \bar{N} 
(v \!\cdot\! u)^2 N 
 + c_3 \bar{N} u \!\cdot\! u  N  
\nonumber \\ 
& - & 
\frac{i}{4M} [S^\mu ,S^\nu ]_{-} 
\left ( (1+\kappa_v) f_{\mu\nu}^+ + \frac{1}{2} 
(\kappa_s - \kappa_v) 
{\mathrm{Tr}}  f_{\mu\nu}^+ \right ) \Big  \} N 
\, + \dots  
\label{lrmc1}  
\end{eqnarray} 
Above we have used standard notations \cite{bkm95}: 
\begin{eqnarray}
D_\mu U \, &\equiv&  \, \partial_\mu U - i ({\cal{V}}_\mu  
+ {\cal{A}}_\mu ) U
+ i U   ({\cal{V}}_\mu  - {\cal{A}}_\mu ) ;
\nonumber \\
U &\equiv& u^2 ;  \hspace{8mm} 
u_\mu \equiv i u^\dagger D_\mu U u^\dagger  
\, = i \, \left(u^\dagger \partial_\mu u - u \partial_\mu u^\dagger \right) ; 
\nonumber \\
D_\mu N \, &\equiv& \, \partial_\mu N + \frac{1}{2} 
[u^\dagger,\partial_\mu u ]_{-} N
- \frac{i}{2} u^\dagger ({\cal{V}}_\mu + {\cal{A}}_\mu ) u 
- \frac{i}{2} u ({\cal{V}}_\mu - {\cal{A}}_\mu ) u^\dagger ; 
\nonumber \\
F_\mu^R &\equiv& {\cal{V}}_\mu + {\cal{A}}_\mu ; \hspace{8mm} 
F_\mu^L \equiv {\cal{V}}_\mu - {\cal{A}}_\mu 
\nonumber \\
F_{\mu\nu}^{L,R} &\equiv& 
\partial_\mu F_\nu^{L,R} - 
\partial_\nu F_\mu^{L,R} -
i [F_\mu^{L,R} , F_\nu^{L,R} ]_-  ; \hspace{3mm} 
f_{\mu\nu}^+ \equiv u^\dagger F_{\mu\nu}^{R} u +
 u F_{\mu\nu}^{L} u^\dagger . 
\label{definitions}
\end{eqnarray} 
The covariant derivatives above include 
the external vector and axial vector fields,
${\cal V}_\mu = {\cal{V}}_\mu^a \frac{\tau^a}{2}$
and ${\cal A}_\mu = {\cal{A}}_\mu^a \frac{\tau^a}{2}$, 
respectively. 
The covariant  spin operator $S^\mu$
of the heavy nucleon 
is  defined by 
$S_\mu \equiv \textstyle 
\frac14 \gamma_5 
[\slash\hskip -0.52em v, \gamma_\mu] $,  
which for our  static velocity  equals 
$S^\mu = (0,\frac{1}{2} \vec{\sigma})$.  
The coupling constants $c_1,c_2$ and $c_3$ 
can be fixed from phenomenology \cite{bkm95}. 
Their values  
are related to the pion-nucleon 
$\sigma$-term, $\sigma_{\pi N}(t) 
\sim \langle p^\prime |
m_q (\bar{u}u + \bar{d}d) |p\rangle$ 
(where $t=(p^\prime - p)^2$), 
the axial polarizability $\alpha_A$, 
and the isospin-even $\pi N$ $s$-wave 
scattering length 
$a^+\equiv \frac13(a_{1/2} + 2 a_{3/2})
\approx -0.008 m_\pi^{-1}$. 
%
The four-Fermi non-derivative contact terms 
in Eq.(\ref{lpin0}) 
were introduced 
by Weinberg \cite{wei90} and 
further investigated in two- and three-nucleon systems 
by van Kolck {\it et al}. \cite{kol92}. 
%
(In Eq.(\ref{lpin0}) the sum over $A$ runs over 
the possible combinations of 
$\gamma$- and $\bbox{\tau}$- matrices: 
$\Gamma_{S}^{S} = 1$, 
$\Gamma_{S}^{V} = \bbox{\tau}$, 
$\Gamma_{V}^{S} = S_\mu$, 
and $\Gamma_{V}^{V} = S_\mu \bbox{\tau}$. 
However, because of the Fermi statistics 
(Fierz rearrangement), 
only two of the four coupling constants $C_A$ 
are independent.) 
Although these terms are important
in the chiral perturbative derivation of
the short range nucleon-nucleon interactions  
not included in multi-pion exchanges 
\cite{wei90,kol92},
they do not play a major role 
in the following discussion.

In addition to the chiral order index $\bar{\nu}$
defined for each term in ${\cal L}_{\rm{ch}}$,
a chiral order index $\nu$ is assigned 
for each irreducible Feynman diagram
appearing in the chiral perturbation series
\cite{wei90}.
Its definition for a multifermion system is 
\begin{equation}
\nu = 4 - E_N - 2C + 2L + \sum_i \bar{\nu}_i,
\label{eq:nununu}
\end{equation}
where $E_N$ is the number of 
nucleons in the Feynman diagram, 
$L$ the number of loops, 
and $C$ the number of disconnected parts
of the diagram.
The sum over $i$ runs over all the vertices 
in the Feynman graph, 
and $\bar{\nu}_i$ is the chiral order 
of each vertex. 
One can show \cite{wei90} that
an irreducible diagram of chiral order $\nu$
carries a factor 
$(\tilde{Q}/\Lambda)^{\nu} \ll 1$. 
However, 
the application of ChPT to nuclei
involves some subtlety. 
As emphasized by Weinberg \cite{wei90},
naive chiral counting
fails for a nucleus, 
which is a loosely bound many-body system.
This is because purely nucleonic intermediate states
occurring in a nucleus
can have very low excitation energies,
which spoils the ordinary chiral counting.
To avoid this difficulty,
one must first classify diagrams
appearing in perturbation series
into irreducible and reducible diagrams, 
according to whether or not a diagram is 
free from purely nucleonic intermediate states.
Thus, in an irreducible diagram,
every intermediate state 
contains at least one meson.
The ChPT can be  applied 
to the irreducible diagrams.
The contribution of 
all the irreducible diagrams 
(up to a specified chiral order)
is then to be used as an effective operator
acting on the nucleonic Hilbert space.
This second step allows us 
to incorporate the contributions of 
the reducible diagrams.
We may refer to this two-step procedure 
as the {\it nuclear chiral perturbation theory} 
(nuclear ChPT).
This method was first applied by Weinberg \cite{wei90} 
and  by van Kolck {\it et al.} \cite{kol92} 
to the few nucleon system. 
Park, Min and Rho \cite{rho} 
applied the nuclear ChPT
to meson exchange currents in nuclei, and others had success 
in describing the exchange currents
for the electromagnetic and weak interactions
\cite{ptk94,pmr95}.

In the literature 
the term ``effective lagrangian"  is 
often used to imply that
the lagrangian 
is only meant for calculating tree diagrams.
We must note, however, that
the ``modern" effective lagrangians 
have a different meaning. 
Not only can ${\cal L}_{\rm{ch}}$ 
of Eq.(\ref{Lag}) be used 
beyond tree approximation
but, in fact, a consistent chiral 
counting demands inclusion of 
every loop diagram whose chiral order $\nu$
is lower than or equal to the chiral order 
of interest. 
As will be discussed below,
for a consistent ChPT treatment of 
the problem at hand,
we therefore need to consider loop corrections 
and necessary lagrangian counterterms of 
chiral order $\nu$ 
to cancel the loop-divergences.


\section{ChPT applications} 
\subsection{Radiative muon capture } 
The radiative muon capture on the proton (RMC) 
$\mu^-+p\rightarrow n+\nu_\mu+\gamma$ 
has an 
extremely small branching ratio  
and to observe RMC 
is a  great experimental challenge.  
Only  recently has an experimental group 
at TRIUMF \cite{jonetal96} finally succeeded
in measuring the 
%
partial capture rate 
$\Gamma_{\mathrm{RMC}}(>60\mathrm{MeV})$, 
corresponding to
emission of a photon with $E_\gamma>60 \mathrm{MeV}$. 
A main goal of measuring 
$\Gamma_{\mathrm{RMC}}$ is to extract 
accurate information 
about the pseudoscalar form factor, $f_P$,
of the weak hadronic matrix element. 
The matrix element of the hadronic charged weak current 
$h^\lambda=V^\lambda -  A^\lambda$
between a proton and a neutron  is 
\begin{eqnarray}
&\langle &
n(p_f) | V^\lambda - A^\lambda | p (p_i) 
\rangle  
\; =  
\nonumber \\
&{\bar {u}}& (p_f) \left[f_V(q^2)\gamma^\lambda
 + \frac{f_M(q^2)}{2m_N}
\sigma^{\lambda\mu}q_\mu
 + f_A(q^2)\gamma^\lambda\gamma_5
 + \frac{f_P(q^2)}{m_\pi}q^\lambda\gamma_5
\right] u(p_i),
\label{eq:formfactors}
\end{eqnarray}
where $q\equiv p_i-p_f$, and 
the absence of the second-class current is assumed. 
Ordinary muon capture on a proton  (OMC), 
$\mu^-+p\rightarrow n+\nu_\mu$,
gives only  an approximate value for $f_P$. 
The reason is that the momentum transfer  in OMC,
$q^2=-0.88 m_\mu^2$, 
is far away from the pion-pole position, $q^2=m_\pi^2$,
where the contribution of $f_P(q^2)$ 
becomes most important. 
RMC 
provides a more sensitive probe of $f_P $  
because in the RMC's three-body final state 
one can be  closer to the pion-pole. 
Using the 
theoretical framework of Beder and Fearing \cite{bf87} 
the TRIUMF group extracted a value for $f_P $ 
about 50\% larger than expected from PCAC. 
This surprising result should be contrasted 
with the fact that $f_P$ measured in OMC
is consistent with the PCAC prediction
within large experimental uncertainties.
In this framework, as in many earlier works
\cite{opa64,fea80},
one invokes a minimal substitution 
to generate the RMC transition amplitude 
from the transition amplitude for OMC,
the hadronic part of which is given by Eq. (\ref{eq:formfactors}).
Essentially,  one writes 
$f_P(q^2)=\tilde {f}_P/(q^2-m_\pi^2)$, 
where $\tilde {f}_P$ is a constant 
and then 
one replaces every $q$ in Eq.(\ref{eq:formfactors})
with $q - e{\cal A}$ 
(${\cal A}$ is the electromagnetic field)
except the $q$ appearing in the $q^2$ dependence 
of $f_V$, $f_A$ and $f_M$. 
Due to the large discrepancy with the expected 
$\tilde{f}_P$ value, it 
seems reasonable to reexamine the reliability
of this existing minimal substitution 
phenomenological approach
\cite{bf87}.

\begin{figure}[ht]
\begin{picture}(380,100)
\ArrowLine(50,0)(70,50)
\ArrowLine(70,50)(50,100)
\ArrowLine(150,0)(130,50)
\ArrowLine(130,50)(150,100)
\ZigZag(70,50)(130,50)28
\Photon(60,25)(25,65)44
\Text(40,0)[c]{$\mu^-$}
\Text(40,100)[c]{$\nu$}
\Text(160,0)[c]{$p$}
\Text(160,100)[c]{$n$}
\Text(15,65)[c]{$\gamma$}
\Text(100,55)[c]{$W^-$}
\Text(100,0)[c]{(a)}
\ArrowLine(250,0)(270,50)
\ArrowLine(270,50)(250,100)
\ArrowLine(350,0)(330,50)
\ArrowLine(330,50)(350,100)
\ZigZag(270,50)(295,50)24
\DashLine(295,50)(330,50)4
\Photon(260,25)(225,65)44
\Text(240,0)[c]{$\mu^-$}
\Text(240,100)[c]{$\nu$}
\Text(360,0)[c]{$p$}
\Text(360,100)[c]{$n$}
\Text(215,65)[c]{$\gamma$}
\Text(285,55)[c]{$W^-$}
\Text(315,55)[c]{$\pi^-$}
\Text(300,0)[c]{(b)}
\end{picture}
\caption{}
\vspace{10mm}
\label{fig1}
\end{figure}
\begin{figure}[ht]
\begin{picture}(380,100)
\ArrowLine(50,0)(70,50)
\ArrowLine(70,50)(50,100)
\ArrowLine(150,0)(130,50)
\ArrowLine(130,50)(150,100)
\ZigZag(70,50)(130,50)28
\Photon(140,25)(175,65)44
\Text(40,0)[c]{$\mu^-$}
\Text(40,100)[c]{$\nu$}
\Text(160,0)[c]{$p$}
\Text(160,100)[c]{$n$}
\Text(185,65)[c]{$\gamma$}
\Text(100,55)[c]{$W^-$}
\Text(100,0)[c]{(a)}
\ArrowLine(250,0)(270,50)
\ArrowLine(270,50)(250,100)
\ArrowLine(350,0)(330,50)
\ArrowLine(330,50)(350,100)
\ZigZag(270,50)(295,50)24
\DashLine(295,50)(330,50)4
\Photon(340,25)(375,65)44
\Text(240,0)[c]{$\mu^-$}
\Text(240,100)[c]{$\nu$}
\Text(360,0)[c]{$p$}
\Text(360,100)[c]{$n$}
\Text(385,65)[c]{$\gamma$}
\Text(285,55)[c]{$W^-$}
\Text(315,55)[c]{$\pi^-$}
\Text(300,0)[c]{(b)}
\end{picture}
\caption{}
\vspace{10mm}
\label{fig2}
\end{figure}
\begin{figure}[ht]
\begin{picture}(380,100)
\ArrowLine(50,0)(70,50)
\ArrowLine(70,50)(50,100)
\ArrowLine(150,0)(130,50)
\ArrowLine(130,50)(150,100)
\ZigZag(70,50)(130,50)28
\Photon(140,75)(185,95)44
\Text(40,0)[c]{$\mu^-$}
\Text(40,100)[c]{$\nu$}
\Text(160,0)[c]{$p$}
\Text(160,100)[c]{$n$}
\Text(195,95)[c]{$\gamma$}
\Text(100,55)[c]{$W^-$}
\Text(100,0)[c]{(a)}
\ArrowLine(250,0)(270,50)
\ArrowLine(270,50)(250,100)
\ArrowLine(350,0)(330,50)
\ArrowLine(330,50)(350,100)
\ZigZag(270,50)(295,50)24
\DashLine(295,50)(330,50)4
\Photon(340,75)(385,95)44
\Text(240,0)[c]{$\mu^-$}
\Text(240,100)[c]{$\nu$}
\Text(360,0)[c]{$p$}
\Text(360,100)[c]{$n$}
\Text(395,95)[c]{$\gamma$}
\Text(285,55)[c]{$W^-$}
\Text(315,55)[c]{$\pi^-$}
\Text(300,0)[c]{(b)}
\end{picture}
\caption{}
\vspace{10mm}
\label{fig3}
\end{figure}

\begin{figure}[ht]
\begin{picture}(380,100)
\ArrowLine(150,0)(170,50)
\ArrowLine(170,50)(150,100)
\ArrowLine(250,0)(230,50)
\ArrowLine(230,50)(250,100)
\ZigZag(170,50)(195,50)24
\DashLine(195,50)(230,50)4
\Photon(220,50)(220,100)44
\Text(140,0)[c]{$\mu^-$}
\Text(140,100)[c]{$\nu$}
\Text(260,0)[c]{$p$}
\Text(260,100)[c]{$n$}
\Text(230,100)[c]{$\gamma$}
\Text(185,35)[c]{$W^-$}
\Text(215,40)[c]{$\pi^-$}
\end{picture}
\caption{}
\vspace{10mm}
\label{fig4}
\end{figure}

\begin{figure}[ht]
\begin{picture}(380,100)
\ArrowLine(50,0)(70,50)
\ArrowLine(70,50)(50,100)
\ArrowLine(150,0)(130,50)
\ArrowLine(130,50)(150,100)
\ZigZag(70,50)(130,50)28
\Photon(130,50)(175,85)44
\Text(40,0)[c]{$\mu^-$}
\Text(40,100)[c]{$\nu$}
\Text(160,0)[c]{$p$}
\Text(160,100)[c]{$n$}
\Text(180,80)[c]{$\gamma$}
\Text(100,55)[c]{$W^-$}
\Text(100,0)[c]{(a)}
\ArrowLine(250,0)(270,50)
\ArrowLine(270,50)(250,100)
\ArrowLine(350,0)(330,50)
\ArrowLine(330,50)(350,100)
\ZigZag(270,50)(295,50)24
\DashLine(295,50)(330,50)4
\Photon(330,50)(375,85)44
\Text(240,0)[c]{$\mu^-$}
\Text(240,100)[c]{$\nu$}
\Text(360,0)[c]{$p$}
\Text(360,100)[c]{$n$}
\Text(380,85)[c]{$\gamma$}
\Text(285,55)[c]{$W^-$}
\Text(315,55)[c]{$\pi^-$}
\Text(300,0)[c]{(b)}
\end{picture}
\caption{}
\vspace{10mm}
\label{fig5}
\end{figure}

\begin{figure}[ht]
\begin{picture}(380,100)
\ArrowLine(150,0)(170,50)
\ArrowLine(170,50)(150,100)
\ArrowLine(250,0)(230,50)
\ArrowLine(230,50)(250,100)
\ZigZag(170,50)(195,50)24
\DashLine(195,50)(230,50)4
\Photon(195,50)(195,100)44
\Text(140,0)[c]{$\mu^-$}
\Text(140,100)[c]{$\nu$}
\Text(260,0)[c]{$p$}
\Text(260,100)[c]{$n$}
\Text(205,100)[c]{$\gamma$}
\Text(185,35)[c]{$W^-$}
\Text(215,40)[c]{$\pi^-$}
\end{picture}
\caption{}
\vspace{10mm}
\label{fig6}
\end{figure} 

ChPT  provides 
a systematic framework to describe
the electromagnetic-, weak-, and strong-interaction
vertices in a consistent manner,
avoiding  
the phenomenological minimal-coupling substitution
at the level of the transition amplitude. 
Furthermore, ChPT also satisfies 
the gauge-invariance and chiral-symmetry requirements
in a transparent way. 
In the case of OMC, Bernard et al.\cite{bkm94} 
and Fearing et al.\cite{feaetal97}
used heavy-baryon ChPT to evaluate 
the value of $\tilde{f}_P$ 
with better accuracy 
but consistent with the value achieved 
in the PCAC approach. 
Muon capture is a favorable case for applying ChPT
since momentum transfers involved here do not exceed $m_\mu$, 
and $m_\mu$ is small compared to the chiral scale 
$\Lambda \sim $ 1 GeV, implying a reasonably 
rapid convergence of chiral expansion.

A  first calculation of  
the total capture rate $\Gamma_{\mathrm{RMC}}$
and the spectrum of the emitted photons, 
$d\Gamma_{\mathrm{RMC}} (k) / dk  $,
in chiral perturbation expansion 
will be briefly discussed \cite{mmk97}. 
We limit ourselves here to 
a next-to-leading chiral order (NLO) calculation
and therefore we only consider the terms 
with ${\bar{\nu}} = 0$ and ${\bar{\nu}} = 1$ 
in Eq.(\ref{Lag}). 
To this chiral order we need only consider tree diagrams,
and then ${\cal L}_{\pi N}^{(1)}$ only represents
$1/M$ ``nucleon recoil'' corrections to 
the leading ``static'' part ${\cal L}_{\pi N}^{(0)}$. 
Furthermore, we do not consider explicit $\Delta$ degrees of freedom. 
The covariant derivatives of Eq.(\ref{definitions}) include 
the external fields,
${\cal V}_\mu $ 
and ${\cal A}_\mu $. 
These are determined by 
the $W^-$ boson which couples to the leptonic 
and hadronic currents in the standard manner. 
In the actual calculation, 
taking the limit $m_W \to \infty$,
we make the substitution 
$ W_\mu^- \to ({\cal{V}}_\mu^- - {\cal{A}}_\mu^-) 
\frac{\tau^1 - i \tau^2}{2} $, 
i.e. we  treat $\cal{V}$ and $\cal{A}$
as static external  sources. 
The only parameters appearing in the 
expressions for 
Eq.(\ref{Lag})  
relevant for RMC 
[Eq.(\ref{lpi0}), first term of Eq.(\ref{lpin0}) and 
first and last line of Eq.(\ref{lrmc1})] 
are the pion decay constant, 
$f_\pi$ = 93 ${\mathrm{MeV}}$, 
the axial vector coupling, $g_A = 1.26$, 
and the nucleon isoscalar and isovector 
anomalous magnetic moments, 
$\kappa_s = -0.12$ and $\kappa_v = 3.71$.
Thus,  to the chiral order of our interest 
${\cal L}_{\mathrm{ch}}$ is well determined.

We consider all possible Feynman diagrams 
Figs.\ref{fig1}-\ref{fig6} 
up to chiral order $\nu$ = 1 
which contribute to the process 
$\mu^- + p \to n + \nu + \gamma$.
The zigzag lines in these diagrams represent
the $W^-$ boson. 
For static $W^-$ bosons the diagrams in Figs.\ref{fig1}-\ref{fig6}
reduce to those that would result from
the simple current-current interaction of the $V-A$ form.
The reason for explicitly retaining the $W^-$ boson lines
is to clearly separate the different photon vertices
(see e.g. Fig.\ref{fig6}). 
The leptonic vertices in these Feynman diagrams are 
of course well known.
The hadronic vertices are obtained 
by expanding the ChPT lagrangian
[Eqs. (\ref{Lag}), (\ref{lpi0}), (\ref{lpin0})
and the relevant parts of (\ref{lrmc1})]
in terms of the elementary fields $N$, $\pi$, 
$\cal{V}$ and $\cal{A}$ and their derivatives. 
The evaluation of the transition amplitudes 
corresponding to these Feynman diagrams 
is straightforward. 
We denote by $M_i$ ($i=1\dots 6$) 
the invariant transition amplitudes
corresponding to Fig.(\ref{fig1})-({\ref{fig6}), respectively. 
They are calculated in the Coulomb gauge, 
i.e. $v \cdot \epsilon (\lambda ) = 0$, 
and are given by:
\begin{eqnarray}
M_1 \, &=& \, \epsilon^\beta (\lambda) 
\left [ \bar{u}_\nu (s) \gamma_\tau (1-\gamma_5 ) 
\frac{ {\mu{\mkern -9 mu}{/}} - 
{k{\mkern -9 mu}{/}} + m_\mu } 
{ 2 (k\cdot\mu ) } \gamma_\beta u_\mu ( s^\prime ) \right ] \;
\left [ N_n^\dagger (\sigma)  h_1^\tau N_p (\sigma^\prime ) \right ]
\label{m1} \\
M_i \, &=& \, \left [ \bar{u}_\nu (s) \gamma_\tau 
(1-\gamma_5 ) u_\mu ( s^\prime ) \right ] \;
\left [ N_n^\dagger (\sigma)  h_i^\tau (\lambda) N_p 
(\sigma^\prime ) \right ]\,,  \; i=2,3,4,5,6 
\label{mi} 
\nonumber
\end{eqnarray} 
where $ h_i^\tau$; $i=1, \cdots ,6$ are  the 
hadronic operators given in Ref.\cite{mmk97}. 
In  Eqs.(\ref{m1}) and (\ref{mi})
$\mu$, $\nu$, $p=(E_p,\vec{p})$, 
$n=(E_n,\vec{n})$ and $k=(\omega_k,\vec{k})$ 
are the four-momenta of the muon, neutrino, 
proton, neutron and photon, respectively. 
The $z$-components of the spins 
of the muon, neutrino, 
proton and neutron are denoted by 
$s$, $s^\prime$, $\sigma^\prime$ and $\sigma$, 
respectively, while $\epsilon(\lambda)$ 
stands for the photon polarization vector.

We consider only the case
of RMC from the $\mu$-$p$ atom 
with statistical spin distributions,
leaving out the hyperfine-state decomposition
and the treatment of RMC from the $p\mu p$ molecule.
With our kinematical approximations:  
both the muon and the proton are at rest 
and neglecting the recoil neutron kinetic energy, 
the spin-averaged total capture rate is  
\begin{eqnarray}
\Gamma_{\mathrm{RMC}} \,  = \,  &&
\left ( 
\frac{eG}{\sqrt{2}} \right )^2
\frac{\vert \Phi (0) \vert^2}{4} 
\,
(2\pi)^4
\,
\int \frac{d^3 n}{(2\pi)^3}
\,
\int \frac{d^3 \nu}{(2\pi)^3}
\,
\int \frac{d^3 k}{(2\pi)^3}
\,
\frac{1}{2 \omega_k}
\nonumber \\ 
&& \times
\delta^{(4)} (n+\nu +k - p - \mu)
\sum_{\sigma{\sigma^\prime}s{s^\prime}\lambda} 
\vert M \vert^2
\label{rmc1}
\end{eqnarray}
where the sum is over all spin and polarization orientations, 
$M = \sum_{i=1}^6 M_i $, 
and the $\mu p$ atomic wavefunction 
at the origin is 
$ \Phi (0) $.  
Within  the kinematical  approximation stated earlier,
$ \delta^{(4)} (n+\nu +k-p-\mu )
\approx \delta( \vert \vec{k} \vert +  
\vert \vec{\nu} \vert - m_\mu ) \delta^{(3)} 
(\vec{n} + \vec{\nu} + \vec{k} ) $ 
and the maximal $\gamma$ energy is 
$(\omega_k)_{\mathrm{max}} \approx m_\mu$. 
The numerical values for the total capture rate   
$\Gamma_{\mathrm{RMC}}$ are then \cite{mmk97}:  \\ 
{\bf 1.} To lowest chiral order $\nu$ = 0 
$\Gamma_{RMC}$ = 0.061 $s^{-1}$ 
(excluding the pion-pole diagrams' 
contributions $\Gamma_{RMC}$ = 0.043 $s^{-1}$). \\ 
{\bf 2.} Including the $\nu$ = 1 (${\cal O} (1/M)$ recoil terms) 
contributions give $\Gamma_{RMC}$ =
0.075 $s^{-1}$ ( no pion-pole = 0.053 $s^{-1}$) \\ 
Our total capture rate 
$\Gamma_{\mathrm{RMC}} = 0.075 s^{-1}$
is close to the value given in \cite{opa64}, 
$\Gamma_{\mathrm{RMC}}\,=\,0.069 s^{-1}$, 
and practically identical to 
$\Gamma_{\mathrm{RMC}}\,=\,0.076 s^{-1}$
reported in \cite{fea80}.
Our  ${\cal{O}} (1/M)$ recoil corrections 
account for about $20\%$ 
of the leading order 
${\cal{O}} ((1/M)^0)$ contribution,
which indicates a reasonable convergence 
of the chiral expansion.      
As one can see  about $30\%$ 
of the total value of $\Gamma_{\mathrm{RMC}}$ 
comes from the pion-pole exchange diagrams 
in our calculation.

A direct comparison of our calculation
with the experimental data \cite{jonetal96}
is premature because we have not considered captures 
from the singlet and triplet hyperfine states separately,
or capture from the $p\mu p$ molecular state.
This also means that at this stage we cannot 
directly address the ``$f_P$ problem'' that arose 
from the TRIUMF data \cite{jonetal96}.
However, 
as already stated ChPT gives a unique prediction on $f_P$, 
are consistent with the Goldberger-Treiman value 
$ f_P = 6.6 g_A $ \cite{bkm94,feaetal97}. 
Meanwhile,
for the spin-averaged $\mu p$-atomic RMC,
our ChPT calculation gives a photon spectrum 
that is harder than that of \cite{bf87} 
for the PCAC value of $f_P$.
Of course, a more quantitative statement can be made 
only after a more detailed ChPT calculation becomes available 
in which the hyperfine states are separated and 
the $p\mu p$-molecular absorption is evaluated.
We must also emphasize that the present calculation
includes only up to the next-to-leading chiral order (NLO) contributions.
A next-to-next-to-leading order (NNLO) calculation
includes the $\nu=2$ chiral lagrangian,
${\cal{L}}_{\pi}^{(2)}$ and ${\cal{L}}_{\pi N}^{(2)}$,
and also loop corrections arising from ${\cal{L}}_{\pi}^{(0)}$ 
and ${\cal{L}}_{\pi N}^{(0)}$ 
has been completed and 
these next order terms are found to give  small 
corrections to the tree diagrams \cite{am97}. 
(The finite contributions from the loop diagrams 
would give momentum-dependent vertices,
which would correspond to the form factors
in the language of the phenomenological approach \cite{bf87,fea80}.) 
We  note that
the approach of Bernard et al.\cite{bkm95}, which we have used here, 
does not contain the explicit $\Delta$ degree of freedom
in contrast to \cite{hhk97}.
For a complete calculation 
the $\Delta$ degrees of freedom 
should be included in the next chiral order 
as done by, e.g.,   
Fearing et al. \cite{feaetal97}.

\subsection{Pion production at threshold } 
The second application of ChPT is concerned with recent 
high-precision measurements 
of the total cross sections near threshold 
for the reaction
$p+p \rightarrow p+p+\pi^0$  
by  Meyer et al. \cite{meyetal90}. 
Traditionally the pion production reactions are 
described by 
the single nucleon process (the Born term), 
Fig.7(a),
and the $s$-wave pion rescattering process,
Fig.7(b). 

\begin{picture}(400,155)


\put(0,40){\line(1,0){25}}
\put(45,40){\line(1,0){70}}
\put(135,40){\line(1,0){25}}

\put(0,90){\line(1,0){25}}
\put(45,90){\line(1,0){70}}
\put(135,90){\line(1,0){25}}

\put(80,90){\line(5,4){10}}
\put(92,99.6){\line(5,4){10}}
\put(104,109.2){\line(5,4){10}}
\put(116,118.8){\line(5,4){10}}
\put(128,128.4){\line(5,4){10}}

\put(35,65){\oval(20,70)}
\put(125,65){\oval(20,70)}

\put(62,82){\makebox(0,0){$p_2$}}
\put(98,82){\makebox(0,0){$p_2'$}}
\put(62,32){\makebox(0,0){$p_1$}}
\put(98,32){\makebox(0,0){$p_1'$}}
\put(100,116){\makebox(0,0){$q$}}

\put(146,136){\makebox(0,0){$\pi^0$}}
\put(-6,90){\makebox(0,0){$p$}}
\put(166,90){\makebox(0,0){$p$}}
\put(-6,40){\makebox(0,0){$p$}}
\put(166,40){\makebox(0,0){$p$}}

\put(10,82){\makebox(0,0){$\bar{p}_2$}}
\put(150,82){\makebox(0,0){$\bar{p}_2'$}}
\put(10,32){\makebox(0,0){$\bar{p}_1$}}
\put(150,32){\makebox(0,0){$\bar{p}_1'$}}

\put(66,0){\makebox(0,0){fig. 7(a)}}



\put(210,40){\line(1,0){25}}
\put(255,40){\line(1,0){70}}
\put(345,40){\line(1,0){25}}

\put(210,90){\line(1,0){25}}
\put(255,90){\line(1,0){70}}
\put(345,90){\line(1,0){25}}

\put(290,90){\line(5,4){10}}
\put(302,99.6){\line(5,4){10}}
\put(314,109.2){\line(5,4){10}}
\put(326,118.8){\line(5,4){10}}
\put(338,128.4){\line(5,4){10}}

\put(290,40){\line(0,1){7}}
\put(290,50){\line(0,1){7}}
\put(290,60){\line(0,1){7}}
\put(290,70){\line(0,1){7}}
\put(290,80){\line(0,1){7}}

\put(245,65){\oval(20,70)}
\put(335,65){\oval(20,70)}

\put(272,82){\makebox(0,0){$p_2$}}
\put(308,82){\makebox(0,0){$p_2'$}}
\put(272,32){\makebox(0,0){$p_1$}}
\put(308,32){\makebox(0,0){$p_1'$}}
\put(310,116){\makebox(0,0){$q$}}

\put(356,136){\makebox(0,0){$\pi^0$}}
\put(204,90){\makebox(0,0){$p$}}
\put(376,90){\makebox(0,0){$p$}}
\put(204,40){\makebox(0,0){$p$}}
\put(376,40){\makebox(0,0){$p$}}
\put(298,62){\makebox(0,0){$\pi^0$}}

\put(220,82){\makebox(0,0){$\bar{p}_2$}}
\put(360,82){\makebox(0,0){$\bar{p}_2'$}}
\put(220,32){\makebox(0,0){$\bar{p}_1$}}
\put(360,32){\makebox(0,0){$\bar{p}_1'$}}
\put(284,62){\makebox(0,0){$k$}}

\put(276,0){\makebox(0,0){fig. 7(b)}}

\end{picture}


The Born term is 
assumed to be given by the 
pseudovector interaction Hamiltonian  
\begin{equation}
{\cal H}_0 = \frac{g_A}{2 f_\pi} \bar{\psi} 
\left( \bbox{\sigma} \!\cdot\! \bbox{\nabla} 
( \bbox{\tau} \!\cdot\! \bbox{\pi} )
- \frac{i}{2 M } 
\{ \bbox{\sigma} \!\cdot\! \bbox{\nabla}, 
\bbox{\tau} \!\cdot\! \dot{\bbox{\pi}} \} 
\right) \psi, 
\label{H0}
\end{equation}
where $g_A$ is the axial coupling constant,
and $f_\pi$ = 93 MeV is 
the pion decay constant.
The first term represents 
$p$-wave pion-nucleon coupling,
while the second term accounts 
for the nucleon recoil effect 
and s-wave pion coupling.  
The s-wave rescattering vertex in Fig.7(b)
is commonly calculated 
using the phenomenological Hamiltonian \cite{kr66} 
\begin{equation}
{\cal H}_{1} = 
4\pi \frac{\lambda_1}{m_\pi} \bar{\psi} 
\bbox{\pi}\!\cdot\!\bbox{\pi} \psi
+ 4\pi \frac{\lambda_2}{m^2_\pi} 
\bar{\psi} \bbox{\tau}\!\cdot\!
\bbox{\pi} \!\times\! \dot{\bbox{\pi}} \psi
\label{H1}
\end{equation}
The two coupling constants 
$\lambda_1$ and $\lambda_2$ determined 
from s-wave 
pion nucleon scattering lengths have the values 
$\lambda_1 \sim 0.005$ and
$\lambda_2 \sim 0.05$. 
Thus, $\lambda_1 \ll \lambda_2$ 
as expected from current algebra. 
The early calculations \cite{kr66,ms91,nis92} 
which are  based on these phenomenological vertices, 
underestimate 
the  measured $\pi^0$ production 
cross sections by a factor of $\sim$5.
Since  the second term in Eq.(\ref{H0}) 
is suppressed compared to the first term 
by $\sim m_\pi / M$, 
the importance of the rescattering term 
is enhanced. 
However,   
$\lambda_2$ is much larger than $\lambda_1$, and 
the isospin structure of the $\lambda_2$ term 
is such that it cannot contribute 
to the $\pi^0$ production 
from two protons at the rescattering vertex in Fig.7(b). 
Thus, the use of the phenomenological 
Hamiltonians, Eqs.(\ref{H0}) and (\ref{H1}), to calculate
the Born term and the rescattering terms,
gives significantly suppressed cross sections for 
the $pp \rightarrow pp\pi^0$ reaction
near threshold compared to $\pi^+$ production 
because only the small correction terms contributes. 
Therefore, the  calculated  cross sections 
can be highly sensitive to 
any deviations from this conventional treatment.

It is convenient for our discussion to introduce the 
{\it typical threshold kinematics}  
for this reaction. 
Consider Fig.7(b) in the center-of-mass  system
with the initial and final interactions turned off 
(since it will modify what follows). 
At threshold,
$(q_0, \bbox{q})$  = $(m_\pi, 0) $, 
$p^\prime_{10} = p^\prime_{20} = M$,  and 
$\bbox{p}_1'=\bbox{p}_2'=0$,
so that any exchanged particle must have
$k_0 = m_\pi/2= 70$ MeV and 
$|\bbox{k}| = \sqrt{m_\pi M+ (m_\pi /2)^2} \sim 370$ MeV/c,
which implies
$k^2 = -m_\pi M $.
Thus the rescattering process 
probes internucleon   distances $\sim$ 0.5 fm.  
The reaction is therefore sensitive to exchange of 
heavy mesons which are important in 
phenomenological $N$-$N$ potentials. 
Lee and Riska\cite{lr93}
showed that  heavy-meson exchanges 
(scalar and vector exchange) 
could be capable of 
enhancing the cross section significantly. 
Meanwhile, Hern\'{a}ndez and Oset \cite{ho95} 
and Hanhart {\it et al.}\cite{hanetal95}
showed that the {\em off-shell} 
dependence of the $\pi N$ $s$-wave isoscalar 
amplitude featuring 
in the rescattering process, 
$k^2 = - m_\pi M \ne m_\pi^2$, 
could also appreciably enhance 
the rescattering amplitude.  
Given these developments 
we consider it important to examine 
the significance of 
these phenomenological lagrangians
in ChPT. 

Systematic studies based on ChPT 
would be valuable to sharpen our 
conclusions on whether or not the 
heavy-meson exchange 
contributions are needed to explain 
the observed cross section for 
$p p \rightarrow p p \pi^0$. 
Three very similar ChPT investigations have 
recently been completed 
\cite{cfmv95,pmmmk,slmk}. 
Here we describe briefly the work of the USC group. 
To generate the one- and two-body diagrams 
of Figs. 7(a) and 7(b) we minimally 
need terms with $\bar{\nu} = $ 0 and 1. 
Eq.(\ref{Lag}) 
leads to the pion-nucleon interaction Hamiltonian
${\cal H}_{int} = {\cal H}^{(0)}_{int} 
+ {\cal H}^{(1)}_{int}$, 
with
\begin{mathletters}
\label{eq:17}
\begin{equation}
 {\cal H}^{(0)}_{int} = 
\frac{g_A}{2f_\pi} \bar{N} 
[ \bbox{\sigma}\!\cdot\!\bbox{\nabla} 
 ( \bbox{\tau}\!\cdot\!\bbox{\pi} ) ] N 
 + \frac{1}{4f_\pi^2}
\bar{N} \bbox{\tau}\!\cdot\!\bbox{\pi}
\!\times\!\dot{\bbox{\pi}} N
\label{eq:Hint0}
\end{equation}
\begin{eqnarray}
{\cal H}^{(1)}_{int} &=&
 \frac{-i g_A}{4m_N f_\pi} \bar{N} 
\{ \bbox{\sigma}\!\cdot\!\bbox{\nabla}, 
\bbox{\tau}\!\cdot\!\dot{\bbox{\pi}} \} N  
+ \frac{1}{f^2_\pi} [ 2c_1 m_\pi^2 \pi^2 
 \!-\! (c_2 \!-\! \frac{g^2_A}{8m_N}) \dot{\pi}^2
 \!-\! c_3 (\partial \pi)^2 ] \bar{N} N 
\label{eq:Hint1}
\end{eqnarray}
\end{mathletters} 
Here ${\cal H}^{(\bar{\nu})}_{int}$
represents the term of chiral order
$\bar{\nu}$.
We now compare ${\cal H}_{int}$ 
resulting from  ChPT, Eq.(\ref{eq:17})
with the phenomenological effective Hamiltonian
${\cal H}_0+{\cal H}_1$,
Eqs.(\ref{H0}) and (\ref{H1}).
We note that the first term in ${\cal H}^{(0)}$
and the first term in ${\cal H}^{(1)}$
exactly reproduces ${\cal H}_0$.
Thus the so-called ``Galilean-invariance" term
naturally arises as a 
$1/M$ correction term in HFF.
As for the $\pi\pi NN$ vertex,
we can associate the second term in $ {\cal H}^{(0)}_{int}$
to the $\lambda_2$ term in ${\cal H}_1$,
and second term in $ {\cal H}^{(1)}_{int}$
to the $\lambda_1$ term in ${\cal H}_1$.
Consequently,  
\begin{equation}
\begin{array}{l}
4\pi \lambda^{\chi}_1/m_\pi 
\equiv 
\frac{m_\pi^2}{f_\pi^2} 
[2c_1 - (c_2 - \frac{g_A^2}{8m_N})
\frac{\omega_q \omega_k}{m_\pi^2} 
 - c_3 \frac{q\cdot k}{m_\pi^2} ] 
\equiv \kappa(k,q)
\end{array}
\label{eq:kappakq}
\end{equation} 
with 
$q=(\omega_q, {\bf q})$ and 
$k=(\omega_k, {\bf k})$. 
Now, for {\it on-shell} low energy pion-nucleon scattering,
{\it i.e.\/}, $k \!\sim\! q \!\sim\! 
(m_\pi, \bbox{0})$, we equate
\renewcommand{\theequation}
{\arabic{equation}}
\begin{equation}
4\pi\lambda^{\chi}_1/m_\pi
= \kappa_0\,\equiv\,
\kappa(k\!=\!(m_\pi, \bbox{0}),
q\!=\!(m_\pi, \bbox{0})),
\label{eq:onshell}
\end{equation}
where
\begin{eqnarray}
\kappa_0&=&\frac{m_\pi^2}{f_\pi^2}
\left( 2c_1-c_2-c_3 + 
\frac{g_A^2}{8 M } \right) 
=\,-2\pi
\left( 1+\frac{m_\pi}{m_N} \right)a^+ +
\frac{3g_A^2}{128\pi}
\frac{m_\pi^3}{f_\pi^4}. 
\label{eq:kappa} 
\end{eqnarray}
The above cited empirical value for $a^+$
leads to 
$\kappa_0 = (0.87\pm0.20)\,{\rm GeV}^{-1}$.
If we keep only the lowest chiral order, 
the first  term in Eq.(\ref{eq:kappa}),  
$\lambda^\chi_1$ is 
\begin{equation}
\frac{4\pi\lambda^\chi_1}{m_\pi}
\,=\,-2\pi\left( 1+\frac{m_\pi}{m_N} 
\right)a^+ = \, 
(0.43\pm0.20)\,{\rm GeV}^{-1}, 
\label{eqll}
\end{equation}
or $\lambda^\chi_1\,=\,0.005\pm0.002$, 
identical to the ``standard value" of 
$\lambda_1$ used in the literature, Eq.(\ref{H1}). 
We note that the ChPT value for $\kappa_0 = 0.87 $ 
GeV$^{-1}$, Eq.(\ref{eq:kappa}), 
include next chiral order terms $\propto g_A^2$, 
and is twice the value of 
the first term, Eq.(\ref{eqll}). 
In our comparison  between the traditional and the
ChPT approaches below, we shall use 
Eq.(\ref{eq:kappakq}).
Obviously, since  $\kappa(k,q)$
depends on the four-momenta
$q$ and $k$, we 
cannot identify $\kappa$ with 
the constant $\lambda_1$. 
To  illustrate how the $q$ and $k$ 
dependencies in $\kappa(k,q)$ 
affect the rescattering amplitude 
of Fig.7(b), we 
consider  the 
{\it typical threshold kinematics}, 
$q \! \sim \! (m_\pi,\bbox{0})$ and 
$k \!\sim\! (\frac12 m_\pi, \sqrt{m_\pi M})$, 
and  denote this $\kappa(k,q)$ value 
by $\kappa_{th}$. 
\begin{equation}
\kappa_{th}=\frac{m_\pi^2}{f_\pi^2} 
\left[ 2c_1 - \frac{1}{2}
\left(c_2 - \frac{g^2_A}{8m_N}\right) -\frac{c_3}{2} 
\right] 
\sim -1.5 {\rm GeV}^{-1}. 
\label{eq:kappath}
\end{equation}
Thus  the 
$s$-wave pion-nucleon interaction 
is {\it not only} much stronger than the on-shell cases,
of Eqs.(\ref{eq:kappa}) and (\ref{eqll}),
{\it but} the sign of the off-shell coupling strength 
is {\it opposite} to the on-shell cases.
The first feature is qualitatively in line with 
the observation of 
Refs.\cite{ho95,hanetal95}
that the rescattering term should be larger than 
previously considered.
However, the sign of the typical off-shell coupling, 
$\kappa_{th}$, is
opposite to the one used in Refs.\cite{ho95,hanetal95}. 
This flip of the sign 
drastically changes the pattern of interplay between the Born 
and rescattering terms. 
The sign change in ChPT arises 
from the $\pi N$ rescattering vertex 
where the $\pi N$ {\it scattering occurs at an 
energy below the $\pi N$ threshold}. 
If we force the exchanged pion on the mass-shell 
and only consider the exchanged pion 
three-momentum $\vert \bbox{k} \vert$ 
to be off-shell, we essentially 
recover the results of Hanhart {\it et al.}
\cite{hanetal95}.

The two-nucleon transition matrix element $T$
for the $pp \rightarrow pp\pi^0$ process
is 
$T\,=\,\langle \Phi_f | {\cal T} | \Phi_i \rangle$,
where $ | \Phi_i \rangle$ ($| \Phi_f \rangle$) 
is the initial (final) two-nucleon state distorted 
by the initial-state 
(final-state) interactions. 
Here ${\cal T}$ represent the contributions of
all irreducible diagrams 
(up to a specified chiral order $\nu$)
for the $pp \rightarrow pp\pi^0$ process, 
and we use ${\cal T}$ 
as an effective transition operator 
in the Hilbert space of nuclear wavefunctions. 
\begin{mathletters}
\label{eq:35}
\begin{equation}
{\cal T}\,=\, {\cal T}^{(-1)}+{\cal T}^{(1)} 
\, \equiv \, {\cal T}^{\mbox{\scriptsize Born}}_{-1} 
+ {\cal T}^{\mbox{\scriptsize Res}}_{+1}
\label{eq:calTtrunc}
\end{equation} 
\begin{equation}
{\cal T}^{\mbox{\scriptsize Born}}_{-1}
 = \frac{g_A}{4m_N f_{\pi}} \omega_q \sum_{i=1,2} 
\bbox{\sigma}_i \!\cdot\! 
(\bbox{p}^\prime_i + \bbox{p}_i ) {\tau}_i^0, 
\label{Tm1} 
\end{equation}
\begin{equation}
{\cal T}^{\mbox{\scriptsize Res}}_{+1}
 = -\frac{g_A}{f_\pi} \sum_{i=1,2} \kappa(k_i,q) 
\frac{\bbox{\sigma}_i \!\cdot\! \bbox{k}_i \tau_i^0 }
      { k_i^2 - m_\pi^2 + i\varepsilon}
\label{Tp1}
\end{equation}
\end{mathletters}
The operator ${\cal T}^{(\nu)}$ represents
the contribution from Feynman diagrams of chiral order $\nu$. 
The initial and final three-
momenta of the $i$-th proton are 
$\bbox{p}_i$ and $\bbox{p^\prime}_i$, 
$ k_i \equiv p_i - p^\prime_i$;
and $\kappa(k_i,q)$ is as defined 
in Eq.(\ref{eq:kappakq}).
To chiral order $\nu=1$ we have 
additional contributions due to the loop corrections. 
These loop corrections to the Born diagram, Fig.7(a) 
introduce an effective form factor at the vertex 
defined by ${\cal L}_{ch}$. 
The loop corrections do not change 
drastically our results. Since we cannot 
reproduce the measured cross section, 
we ignore these corrections in this paper.

A formally ``consistent" treatment of $T$ 
would consist in using for $|\Phi_i\rangle$ 
and $|\Phi_f\rangle$ 
two-nucleon wave functions 
generated by irreducible diagrams
of order up to $\nu=1$. 
A problem in this ``consistent" ChPT approach is
that the intermediate two-nucleon propagators 
in Fig.7 can be significantly off-mass-shell. 
Another more practical problem is 
that, if we include the initial- and final- 
two-nucleon ($N$-$N$) interactions
in diagrams up to chiral order $\nu=1$, 
these $N$-$N$ interactions are not 
realistic enough to reproduce the known $N$-$N$ observables.
A pragmatic remedy for these problems is to use 
a phenomenological $N$-$N$ potential
to generate the distorted $N$-$N$ wavefunctions. 
Park, Min and Rho \cite{pmr95} used 
this hybrid approach to study the exchange-current
in the $n+p \rightarrow \gamma+d$ reaction and 
at least, for the low-momentum transfer process
studied in Ref.\cite{pmr95},
the hybrid method is known to work extremely well.

Following standard practice in nuclear physics 
the $T$-matrix was evaluated using $r$-space 
wave functions after Fourier transforming 
${\cal T}^{\mbox{\scriptsize Born}}_{-1}$ and 
${\cal T}^{\mbox{\scriptsize Res}}_{+1}$,
Eq.(\ref{eq:35}),  into $r$-space. 
To simplify the Fourier transformation the 
{\it fixed kinematics approximation}, 
$\kappa(k,q)$  = $\kappa_{th}$, Eq.(\ref{eq:kappath}), 
was used\cite{cfmv95,pmmmk}. 
The results of these two groups  indicate that,
for the various nuclear distortion 
potentials considered, 
the calculated cross section is 
much too small 
to reproduce the experimental cross section. 
If we define the discrepancy ratio $R$ by
$R \equiv \sigma_{tot}^{exp}/\sigma_{tot}^{calc} $,
with $\sigma_{tot}^{exp}$ taken from Ref.\cite{meyetal90},
then $R\,\cong\,80$ ($R\,\cong\,210$)
for the Hamada-Johnston (Reid soft-core) potential, 
and $R$ happens to be almost constant
for the whole range of $E_f\leq 23$ MeV for which
$\sigma_{tot}^{exp}$ is known \cite{pmmmk}. 
Thus, the use of  $\kappa_{th}$, 
Eq.(\ref{eq:kappath}),
results in a significant cancellation 
between the Born- 
and the rescattering terms. 
This destructive interference is in sharp 
contrast to the constructive interference of 
Refs.\cite{kr66,lr93,ho95,hanetal95}.
However, using ChPT in a systematic fashion 
we have shown that the contribution of 
the pion rescattering term 
can be much larger than obtained in the traditional 
phenomenological calculations.

We also learn that the {\it fixed kinematics approximation} 
(which is commonly used in the literature)
should be avoided. There are 
at least two reasons why this is not a good
approximation for this reaction:
(i) The initial- and final-state interactions
play an essential role in the near-threshold 
$pp \rightarrow pp\pi^0$ reaction;
(ii) The theoretical cross section
within the framework of the Born plus 
rescattering terms is likely to depend on
the delicate cancellation between these two terms.
In a just completed momentum space calculation 
we avoid the $\kappa (k,q)$ 
= $\kappa_{th}$ and the 
{\it fixed kinematics approximation}\cite{slmk}.  
We use Eq.(\ref{eq:35}) directly where 
$\kappa (k,q)$ is given by Eq.(\ref{eq:kappakq}). 
Our results show that the magnitude of 
${\cal T}^{\mbox{\scriptsize Res}}_{+1}$ 
increases by a factor $\sim 3$ which means 
the $\pi N$ rescattering term, Fig.7(b) dominates 
\cite{slmk}. 
Further, we confirm the sign of the 
rescattering amplitude found in the 
first two  ChPT calculations \cite{cfmv95,pmmmk}, and 
we find that the $c_1$ term in 
$\kappa (k,q)$  dominates since the $\pi N$ rescattering 
terms 
$c_2$ and $c_3$, Eq.(\ref{eq:kappakq})  
average to a tiny value  in the 
distorted wave integrals due to their energy dependences. 
The constant $c_1$ is given by the $\pi N$ sigma term, 
$\Sigma_{\pi N} (0)$, and the 
uncertainty in the magnitude of the rescattering amplitude, 
$ T^{\mbox{\scriptsize Res}}_{+1}$, 
is given directly by the uncertainty in the value of 
$\Sigma_{\pi N} (0)$. 
Due to the destructive interference with the Born term, 
this uncertainty is enhanced in the resulting cross section. 
However, our 
calculated cross section is still too small. 
Presently we \cite{dkms} are evaluating the next chiral order 
two-pion exchange 
diagrams which 
simulate partially the $\sigma$-meson exchange 
used in Ref.\cite{lr93}. 
To achieve a large cross section these terms 
should be important and this call into question 
the convergence of ChPT.  
Phenomenologically one could also include the repulsive 
$\omega$-exchange when we compare with experiments. 
In the effective ${\cal L}_{ch}$ of Eq.(\ref{lpin0}) 
this is included in the four-nucleon terms, where we would 
calculate the constant $C_A$ as given by $\omega$ exchange. 
For charged pion production from $p p$, 
the higher chiral order 
terms considered here are smaller corrections 
to the dominant Born- and rescattering terms. 
For $\pi^0$ production, however, 
these next order chiral correction terms 
give  the dominant amplitudes.

Finally, we should mention that 
when ChPT is extended to SU$_F$(3) 
and we assume the $K N$ sigma term 
$\Sigma_{KN} (0)$  $\sim c_1$ to dominate, 
the effective Kaon mass in nuclear matter could become very small 
and we could have a Kaon-condensate in dense nuclear matter. 
However, if we use current algebra, ``weak" PCAC 
and standard nuclear physics treatment of meson self-energy in matter, 
we do not find a Kaon condensate in matter. 
This is still an open question 
and the existence of a Kaon condensate 
in dense nuclear matter opens exciting possibilities 
in astrophysics, e.g. 
Ref.\cite{bb94}. 

\nopagebreak
\hspace*{2.25in}{\bf ACKNOWLEDGEMENTS} 

This work is supported in part 
by the National Science Foundation, 
Grant No. PHYS-9602000.

\nopagebreak



\begin{thebibliography}{99}

\bibitem{bb94} G.E. Brown and H.A. Bethe, 
Astrophys. J. {\bf 423} (1994) 659. 

\bibitem{gl84}
J. Gasser and H. Leutwyler, 
Ann. Phys. N.Y. {\bf 158} (1984) 142;
Nucl. Phys. {\bf B250} (1985) 465.

\bibitem{ecker}
For a review, see e.g. 
G. Ecker, Prog. Part. Nucl. Phys. {\bf 35} (1995) 1.

\bibitem{bkm95}
For a review, see e.g.
V. Bernard, N. Kaiser and U.-G. Meissner, 
Int. J. Mod. Phys. {\bf E4} (1995) 193.

\bibitem{vankolck}
C. Ordonez, L. Ray and U. van Kolck, Phys. Rev. Lett. {\bf 72} (1994) 1982;
Phys. Rev. C 53, (1996) 2086.

\bibitem{rho}
T.S. Park, D.-P. Min and M. Rho, Phys. Rep. {\bf 233} (1993) 341.

\bibitem{jm91}
E. Jenkins and A. V. Manohar, 
Phys. Lett. {\bf B255} (1991) 558; 
Phys. Lett. {\bf B259} (1991) 353.

\bibitem{wei90} 
S. Weinberg, 
Phys. Lett. {\bf B251}, 288 (1990);
Nucl. Phys. {\bf B363}, 3 (1991);
Phys. Lett. {\bf B295}, 114 (1992).


\bibitem{kol92}
U. van Kolck, thesis, University of Texas at Austin,
(1992);
C. Ordonez, L. Ray and U. van Kolck,
Phys. Rev. Letters, {\bf 72}, 1982 (1994).



\bibitem{ptk94}
T. S. Park, I. S. Towner and K. Kubodera, 
Nucl. Phys. {\bf A579} (1994) 381.

\bibitem{pmr95}
T. S. Park, D.-P. Min and M. Rho, 
Phys. Rev. Lett. {\bf 74} (1995) 4153; 
``Chiral Lagrangian approach to exchange 
vector currents in nuclei",
preprint SNUTP 95-043 (nucl-th/9505017), 1995.


\bibitem{jonetal96}
G. Jonkmans et al., Phys. Rev. Lett. {\bf 77} (1996) 4512.
 
\bibitem{bf87}
D.S. Beder and H.W. Fearing, Phys. Rev. {\bf D 35} (1987) 2130;
Phys. Rev. {\bf D 39} (1989) 3493.


\bibitem{opa64}
G.I. Opat, Phys. Rev. {\bf 134} (1964) B428.


\bibitem{fea80}
H.W. Fearing, Phys. Rev. {\bf C 21} (1980) 1951.



\bibitem{bkm94}
V. Bernard, N. Kaiser and U.-G. Meissner, 
Phys. Rev. {\bf D 50} (1994) 6899.

\bibitem{feaetal97}
H.W. Fearing, R. Lewis, N. Mobed and S. Scherer, preprint: 
hep-ph/9702394. 

\bibitem{mmk97} T. Meissner, F. Myhrer and K. Kubodera, 
USC-preprint, nucl-th/9707019 (1997), 
submitted to Phys. Letters. 

\bibitem{am97} S. Ando and D.-P. Min, preprint: 
hep-ph/970504 (1997).

\bibitem{hhk97}
T. Hemmert, B. Holstein and J. Kambor,
Phys. Lett. {\bf B 395} (1997) 89.

\bibitem{meyetal90}
H. O. Meyer {\it et al.\/}, 
Phys. Rev. Lett. {\bf 65} (1990) 2846;
Nucl. Phys. {\bf A539} (1992) 633.


\bibitem{kr66}
D. S. Koltun and A. Reitan, 
Phys. Rev. {\bf 141} (1966) 1413.

\bibitem{ms91}
G. A. Miller and P. U. Sauer, 
Phys. Rev. C {\bf 44} (1991) R1725.

\bibitem{nis92} 
J.A. Niskanen,  Phys. Lett. {\bf B289} (1992) 227.

\bibitem{lr93}
T.-S. H. Lee and D. O. Riska, 
Phys. Rev. Lett. {\bf 70} (1993) 2237.

\bibitem{ho95}
E. Hern\'{a}ndez and E. Oset, 
Phys. Lett. {\bf B350}  (1995) 158.

\bibitem{hanetal95}
C. Hanhart, J. Haidenbauer, A. Reuber, 
C. Sch\"{u}tz and J. Speth,  Phys. Lett. 
{\bf B358} (1995) 21; Acta Phys. Polon., 
{\bf B27} (1996) 2893. 


\bibitem{cfmv95}
T.D. Cohen, J.L. Friar, G.A. Miller and U. van Kolck,
Phys. Rev. {\bf C53} (1995) 2661. 

\bibitem{pmmmk} B.-Y. Park, F. Myhrer, J.R. Morones, 
T. Meissner and K. Kubodera, Phys. Rev. {\bf C53} (1996) 1519. 

\bibitem{slmk} T. Sato, T.-S. H. Lee, F. Myhrer and K. Kubodera, 
Phys. Rev. {\bf C56} (1997) 1246. 

\bibitem{dkms} V. Dmitra\v{s}inovi\'{c}, 
T. Sato, F. Myhrer and K. Kubodera, 
in progress. 


\end{thebibliography}
\end{document}